\documentstyle[12pt]{article}
\newcommand{\alt}{\mathbin{\lower 3pt\hbox
   {$\rlap{\raise 5pt\hbox{$\char'074$}}\mathchar"7218$}}}
\newcommand{\agt}{\mathbin{\lower 3pt\hbox
   {$\rlap{\raise 5pt\hbox{$\char'076$}}\mathchar"7218$}}}

\begin{document}
\setcounter{footnote}{0}
\setcounter{equation}{0}
\setcounter{figure}{0}
\setcounter{table}{0}
\vspace*{5mm}

\begin{center}
{\large\bf  Anderson Transition and Generalized Lyapunov Exponents \\
(comment on comment by P.Markos, L.Schweitzer and M.Weyrauch,
cond-mat/0402068) }

\vspace{4mm}
I. M. Suslov \\
P.L.Kapitza Institute for Physical Problems,
\\ 119337 Moscow, Russia \\
\vspace{6mm}

\begin{minipage}{135mm}
{\rm {\quad}  The generalized Lyapunov exponents describe the growth
of the second moments for a particular solution of the quasi-1D
Schroedinger equation with initial conditions on the left end.
Their possible application in the Anderson transition theory
became recently a subject for controversy in the literature.
The approach to the problem of the second moments advanced by
Markos et al (cond-mat/0402068) is shown to be trivially incorrect.
The difference of approaches by Kuzovkov et al (cond-mat/0212036, 0501446)
and the present author (cond-mat/0504557, 0512708) is discussed.  }
\end{minipage}
\end{center}

Recently Markos et al have published a comment \cite{0} on the paper
by Kuzovkov et al \cite{1} where a growth of the second moments
for a particular solution of the quasi-1D Schroedinger equation
was related with the problem of Anderson localization. It was
stated in \cite{1} that the Anderson transition is of the first
order and exists not only for  space dimensions $d>2$ \cite{2}
but also in the 2D case. These statements look wild and the authors
of \cite{0} are right in not believing them. They are also right
in statement that the growth of the second moment of wave function
does not mean the growth of its typical value, which is governed by
the average logarithm. However, Markos et al \cite{0} go
further and claim that analysis of the second moments presented
by Kuzovkov et al \cite{1,2} is qualitatively incorrect and
cannot provide any evidence for the metallic phase.
These conclusions contradict the recent papers by the present
author \cite{3,4} and are shown below to be trivially incorrect.
The difference of approaches suggested in \cite{1,2} and
\cite{3,4} is also discussed.

Consider the 2D Anderson model described by the discrete Schroedinger
equation
$$
\psi_{n+1,m}+\psi_{n-1,m}+\psi_{n,m+1}+\psi_{n,m-1}+ V_{n,m} \psi_{n,m} =
E\psi_{n,m} \eqno(1)
$$
and interprete it as a recurrence relation in the variable $n$, which we
accept as a longitudinal coordinate.  Initial conditions are assumed
to be fixed on the left end of the system, while the periodic boundary
conditions are accepted in the transverse direction,
$\psi_{n,m+L}=\psi_{n,m}$.  Cite energies $V_{n,m}$ are
considered as uncorrelated random quantities with the first two moments
$$
\langle\,V_{n,m} \,\rangle=0\,,\qquad
\langle\,V_{n,m} V_{n',m'} \,\rangle = W^2 \delta_{n,n'} \delta_{m,m'} \,.
\eqno(2)
$$
The growth of the second moments for this problem can be studied using the old
idea by Thouless \cite{5} based on the observation that
variables $\psi_{n,m}$ are statistically independent of $V_{n,m}$
with the same $n$. The main quantity of interest is
$\langle \psi_{n,m}^2\rangle$;  solving (1) for
 $\psi_{n+1,m}$ and averaging its square, we can relate
it with the pair correlators containing lower values of $n$.
Deriving analogous equations for the pair
correlators, we end with the closed system of difference equations for the
quantities
$$
x_{m,m'}(n)\equiv \langle\,\psi_{n,m}\psi_{n,m'} \,\rangle\,,
$$
$$
\,\,y_{m,m'}(n)\equiv \langle\,\psi_{n,m}\psi_{n-1,m'} \,\rangle\,,
\eqno(3)
$$
$$
z_{m,m'}(n)\equiv \langle\,\psi_{n-1,m}\psi_{n,m'} \,\rangle \,,
$$
which for $E=0$  has a form \cite{3}
$$
x_{m,m'}(n+1)=W^2 \delta_{m,m'} x_{m,m'}(n) +x_{m+1,m'+1}(n)+
x_{m-1,m'+1}(n)+ x_{m+1,m'-1}(n) +x_{m-1,m'-1}(n) +
$$
$$
 +x_{m,m'}(n-1)
+y_{m+1,m'}(n) +y_{m-1,m'}(n) +z_{m,m'+1}(n)+ z_{m,m'-1}(n)
\eqno(4)
$$
$$
y_{m,m'}(n+1)=- x_{m+1,m'}(n) -x_{m-1,m'}(n)- z_{m,m'}(n)
\phantom{nnnnnnnnnnnnnnnnnnnnnnnnnnnnnnnnnnnnnnnnnnnnnnnnnnmmmmmmmmmmmmmm}
$$
$$
z_{m,m'}(n+1)=- x_{m,m'+1}(n) -x_{m,m'-1}(n)- y_{m,m'}(n)\,.
\phantom{nnnnnnnnnnnnnnnnnnnnnnnnnnnnnnnnnnnnnnnnnnnnnnnnnnmmmmmmmmmmmmmm}
$$
Instead to follow this natural procedure, Markos et al \cite{0}
invent their own approach. They rewrite (1)
using the transfer matrix and construct the tensor product of two
such matrices. After averaging, they arrive to a linear system of
equations determined by a matrix
$$
T=
  \left ( \begin{array}{cccc}
W^2\,1\bigotimes 1+ D_0 \bigotimes D_0 & -D_0\bigotimes 1 &
-1 \bigotimes D_0& 1 \bigotimes 1\\
D_0 \bigotimes 1 &  0 & -1 \bigotimes 1& 0 \\
1 \bigotimes D_0 & -1 \bigotimes 1&  0& 0\\
1 \bigotimes 1& 0& 0& 0
\end{array} \right) \,,
\eqno(5)
$$
where $D_0=E-H_0$ and $H_0$ is the Hamiltonian of the $n$-th slice for
a pure system, $1$ is the unit matrix of the size $L\times L$. Even
if being correct, this system is practically untractable due to
sophisticated matrix constructions. The eigenvalues $\lambda=\exp(iq)$
of the matrix (5) are declared to be determined by equation
$$
2 \cos{2q} - 2\kappa_i \kappa_j \cos{q} + (\kappa_i^2+ \kappa_j^2-2)
=2W^2 i\sin{q}
\eqno(6)
$$
where $\kappa_i=E-2\cos{p_i}$ and $p_i$ are allowed values of the
transverse momentum $p$. Equation (6) has no resemblance with the
corresponding Eq.\,45  in \cite{3}. The main difference is the
absense of functions with the argument $qL$, which inevitably
arise due to the boundary conditions and can be absent only in
trivial cases. One can suspect, that this difference is related
with incorrect treatment of the disorder term (like $W^2 1\times 1$
in (5)), without which the problem is trivial. This term is local
in $m-m'$ (see (4)) and nondiagonal in the momentum representation,
which is seemingly used in (6). It looks likely that the local
nature of this term was neglected and it was replaced by a
suitable constant. We can try such thing for the system (4),
replacing $\delta_{m,m'}$ by unity. Then (4) is solved trivially
$$
x_{m,m'}(n) =x {\rm e}^{\,i p m+i p' m'}\, {\rm e}^{\,\beta n}\,,\qquad
y_{m,m'}(n) =y {\rm e}^{\,i p m+i p' m'}\, {\rm e}^{\,\beta n}\,,\qquad
z_{m,m'}(n) =z {\rm e}^{\,i p m+i p' m'}\, {\rm e}^{\,\beta n}\,,\qquad
\eqno(7)
$$
where allowed values for $p$ and $p'$  ($2 \pi s/L$, $s=0,1,\ldots,L-1$)
are determined by the boundary conditions and the quantities $x$, $y$, $z$
satisfy the equation
$$
  \left ( \begin{array}{ccc}
W^2 +4\cos{p}\cos{p'}-2\sinh{\beta} & 2\cos{p} & 2\cos{p'}\\
2\cos{p} & {\rm e}^\beta & 1  \\
2\cos{p'} &  1 &  {\rm e}^\beta
\end{array} \right) \,
  \left ( \begin{array}{c}
x\\y\\z
\end{array} \right) \, =0  \,.
\eqno(8)
$$
The determinant vanishes under condition
$$
2 \cosh{2\beta} - 2\kappa \kappa' \cosh{\beta} + (\kappa^2+ \kappa'^2-2)
=2W^2 \sinh{\beta}\,
\eqno(9)
$$
where $\kappa=-2\cos{p}$, $\kappa'=-2\cos{p'}$. In the case $E=0$
(when Eq.4 holds) this equation is identical to (6), if correspondence
$\beta=iq$ is taken into account. We see that, indeed, the disorder term
in (5) was treated inadequately and its local nature was neglected.
 Physically, the equations (6,\,\,9) correspond not to the true Anderson model
 but to its degenerate version when cite energies $V_{n,m}$ are independent
 of $m$.

In fact, the error is present already in the matrix (5). To obtain
the term with disorder,  one needs to produce averaging of the kind
$$
\left ( \begin{array}{cc}
V_1 & 0 \\
0 & V_2 \end{array} \right) \,
\bigotimes
\left ( \begin{array}{cc}
V_1 & 0 \\
0 & V_2 \end{array} \right) \,=
\left ( \begin{array}{cccc}
V_1 V_1 & 0 & 0 &0 \\
0 &  V_1 V_2 & 0 & 0 \\
0 & 0 &  V_2 V_1 & 0\\
0& 0& 0& V_2 V_2
\end{array} \right) \,
\longrightarrow W^2
\left ( \begin{array}{cccc}
1 & 0 & 0 &0 \\
0 & 0 & 0 & 0 \\
0 & 0 &  0 & 0\\
0& 0& 0& 1
\end{array} \right) \,
\eqno(10)
$$
and the result cannot be represented as $W^2\,1\bigotimes 1$. The
latter form is valid for $V_1=V_2$ in Eq.10, while in the general
case it corresponds to the model (1) with $V_{n,m}$ being independent
of $m$, in accordance with the previous analysis. Such model is
of no interest and all conclusions made in \cite{0} are irrelevant
for the problem under consideration.

We see that Markos et al \cite{0} became victims of their own formalism
and were unable to make a way through the jungles of the tensor algebra.
\vspace{2mm}

$\qquad \qquad \qquad \qquad \qquad\qquad$-------------------------------------
\vspace{2mm}

Now let us discuss the difference of approaches suggested by Kuzovkov et al
\cite{1,2} and the present author \cite{3,4}. The initial system of
equations (4) and its higher dimensional analogue
(Eq.\,5 in \cite{4}) coincide with those used in \cite{1,2}. However,
the quantity $z_{mm'}(n)$ was not introduced in \cite{1,2} and its
role was played by $y_{m'm}(n)$. As a result, the system of equations
had no complete difference form and could not be solved in the natural
manner with evaluation of full spectrum of exponents $\beta_s$.
The $Z$-transform exploited in \cite{1,2} allow to find a solution
only in the thermodynamical limit $L\to\infty$, using a
questionable procedure of averaging over translations in the transversal
direction. Translational invariance takes no place for the solution
of (4) (see Eq.\,10 in \cite{4}) and the latter procedure probably
eliminates all terms with the transverse momentum different from $\bf p=0$
and $\bf p=G/2$, if effectively only squares of these terms are relevant
($\bf G$ is a vector of a reciprocal lattice
corresponding to the main diagonal of the Brillioun zone). As a result,
a zero value for the critical disorder was obtained for $d=3$ in the band
center $E=0$ \cite{2}, in a striking contrast with \cite{4}. Fortunately for the
authors of \cite{1,2}, a condition $\bf p=G/2$ corresponds to the minimal
exponent $\beta_{min}$  for $d=2$ and $d\ge 4$, so
the critical values $\sigma'_0$
(corresponding to $W_c$ in \cite{3,4}) were found correctly for these cases. As
for the second critical point $\sigma_0$ for higher dimensionalities,  we see
no evidence for it in the spectrum of $\beta_s$. It looks that a filter
function $H(z)$ used in \cite{1,2} has not only poles corresponding to
eigenvalues of the transfer matrix but also another singilarities,
which are physically irrelevant. Correspondingly, we see no
evidence of a special role of dimensionality $d=6$, which is
surely absent in the exact field theory approach \cite{6}.

Relation of the generalized exponents $\beta_s$ with the
Anderson transition was established in \cite{3,4} using
the conventional variant of finite-size scaling  \cite{7}.
Contrary, the papers \cite{1,2} were formulated in the engineer's
language (using the concepts of signals, filters etc.), which has no
direct relation to the Anderson transition. The limit $L\to\infty$
was taken in \cite{1,2} from the very beginning and the finite-size
scaling approach could not be used for interpretation of results.
As a consequence, a transition in the $2D$ case was interpreted
as being of the first order, in evident contradiction with
all available information. In fact, this transition is of the
Kosterlitz-Thouless type \cite{3} and there is no need for a serious
revision in the weak localization region. The possibility of power-law
localization is mentioned in both approaches \cite{1,2} and \cite{3,4},
but in completely different contexts.

Contrary to \cite{1,2}, a clear distinction is made in \cite{3,4}
between the generalized exponents $\beta_s$ and the true
Lyapunov exponents $\gamma_s$: the latter are self-averaging quantities
and surely have a more fundamental character. Fortunately, the knowledge of
$\beta_s$ provides essential information on $\gamma_s$:
(a) inequality $\beta_s\ge 2 \gamma_s$ can be rigorously proven;
(b) the order of magnitude relation $\beta_s\sim \gamma_s$ takes
place on the physical level of rigorousness;
(c) $\beta_s$ and $\gamma_s$ are practically
equivalent from viewpoint of finite-size scaling philosophy.
Relation of $\beta_s$ and $\gamma_s$ with the parameters of
the log-normal distribution is also of great importance.
In fact, inequality $\beta_s\ge 2 \gamma_s$ is sufficient for
the most responsible statements, such as existence of the $2D$ phase
transition and absense of one-parameter scaling for $\gamma_s$.
Relation $\beta_s\sim \gamma_s$  is used with great precaution
and only in the cases when it does not contradict to numerical
results.

Finally, we do not consider as indisputable the conventional
variant of finite-size scaling based on relation of the Anderson
transition with the minimal Lyapunov exponent $\gamma_{min}$.
In the general case, some effective exponent $\gamma_{eff}$
should be used instead $\gamma_{min}$ \cite{3}. Such
modification restores one-parameter scaling in the weak localization
region and eliminates the $2D$ transition from roughly
half of models.
\vspace{3mm}

This work is partially supported by RFBR  (grant 03-02-17519).


\end{document}